\documentclass[aps,prb,reprint,superscriptaddress,longbibliography,floatfix]{revtex4-2}

\usepackage{amsmath}
\usepackage{amssymb}
\usepackage{bm}
\usepackage{graphicx}
\usepackage{microtype}
\usepackage{booktabs}
\usepackage{array}
\graphicspath{{figures/}{./}}

\newcommand{\avg}[1]{\left\langle #1\right\rangle}
\newcommand{\dd}{\mathrm{d}}
\newcommand{\e}{\mathrm{e}}
\newcommand{\ii}{\mathrm{i}}
\newcommand{\sinc}{\operatorname{sinc}}
\newcommand{\arsinh}{\operatorname{arsinh}}

\begin{document}

\title{Curvature-amplified angular localization of radially localized states}

\author{Hiroyuki Shima}
\email{hshima@yamanashi.ac.jp}
\affiliation{Department of Environmental Sciences, University of Yamanashi, 4-4-37 Takeda, Kofu, Yamanashi 400-8510, Japan}

\date{July 31, 2026}

\begin{abstract}
We study a continuum Schr\"odinger particle on a surface of constant curvature subject to a radial random potential and a weaker angle-dependent perturbation. Exact angular-momentum decomposition reduces the rotationally symmetric problem to independent one-dimensional random radial channels. A positive radial Lyapunov exponent distinguishes radial Anderson localization from the circular probability profile imposed by symmetry. Projection of a narrow radial mode produces a disordered one-dimensional ring. For angular disorder that is stationary in physical arc length, the ring localization length grows as the inverse square of the perturbation amplitude. Equating this length with the geodesic circumference yields a shell-breaking radius that grows algebraically in flat space but as $2a\ln(1/\varepsilon)+O(1)$ on a hyperbolic surface of curvature radius $a$ and angular-disorder amplitude $\varepsilon$. Transfer-matrix and finite-ring calculations determine the crossover coefficient. Direct polar-grid calculations at the same angular energy as the ring model use unbiased tracking of the degenerate angular-momentum subspace across six independent radial-disorder realizations and quantify the validity of the single-mode projection. The noncommuting weak-disorder and flat-curvature limits establish a curvature-amplified crossover for this deliberately anisotropic disorder ensemble, rather than a generic metal--insulator transition on the hyperbolic plane.
\end{abstract}

\maketitle

\section{Introduction}
\label{sec:introduction}

Anderson localization results from coherent multiple scattering by quenched disorder and is conventionally classified by spatial dimension and symmetry class~\cite{Anderson1958,Abrahams1979,LeeRamakrishnan1985,KramerMacKinnon1993,EversMirlin2008}. Correlations in one-dimensional disorder can substantially modify localization lengths and, for suitable long-range correlations, produce extended spectral windows~\cite{IzrailevKrokhin1999,ShimaNomuraNakayama2004}. The extraction of thermodynamic localization properties from finite systems also requires particular care when correlations generate large intrinsic length scales~\cite{NishinoYakuboShima2009}. These observations motivate separating genuine asymptotic localization from finite-size crossovers before assigning a phase-transition interpretation.

Negative curvature changes the long-distance geometry underlying localization. On a surface of constant negative Gaussian curvature, the circumference and area of a geodesic disk grow exponentially with its radius. Closely related work has addressed Anderson transitions and mobility edges on regular hyperbolic lattices~\cite{Chen2024PRL,Li2024CommunPhys}, suppression of the ordinary weak-localization divergence on negatively curved continuum surfaces~\cite{Curtis2025PRL}, geometric delocalization on two-dimensional surfaces containing negative-curvature regions~\cite{Shou2025PRL}, and the renormalization flow of generic isotropic continuum disorder on the hyperbolic plane~\cite{Altland2026}. The last problem concerns bulk conductivity and a metal--insulator critical line. The present problem is intentionally different: a dominant radial random potential first creates a localized mode, and a weaker ring-correlated perturbation controls that mode's angular fragmentation. The shell-resolved crossover derived below is therefore neither a replacement for nor a competing prediction to the isotropic-disorder transition of Ref.~\cite{Altland2026}.

Earlier studies of classical spin systems on negatively curved surfaces established that the same exponential growth also produces strong boundary sensitivity, frustration, and unusual finite-size scaling~\cite{ShimaSakaniwa2006JPA,ShimaSakaniwa2006JSTAT,BaekShimaKim2009,SakaniwaShima2009,BaekMinnhagenShimaKim2009}. Those results reinforce the need to isolate bulk physics from the nonvanishing boundary fraction of a finite hyperbolic disk.

Here we examine a deliberately correlated continuum disorder ensemble. Its dominant component is random in the geodesic radial coordinate and constant along each geodesic circle. This rotationally symmetric limit is an exactly reducible reference problem: each angular-momentum sector is a one-dimensional random radial Hamiltonian. The circular shape of a probability distribution in this limit is therefore imposed by symmetry and is not, by itself, a new localization mechanism. The nontrivial ingredients are (i) a positive radial Lyapunov exponent generated by disorder and (ii) the response of the radially localized state to a weak angle-dependent random perturbation.

Projection onto one radial mode converts the angular problem into a one-dimensional disordered ring whose physical circumference is determined by the metric. The circumference grows linearly with radius in the Euclidean plane but exponentially on the hyperbolic plane. Consequently, the distance at which an initially rotationally extended shell becomes angularly localized is algebraic in the inverse perturbation amplitude in flat space and logarithmic on a negatively curved surface. This geometric conversion is the central result of the paper.

The analysis combines complementary checks. Clean and periodic controls separate a symmetry-enforced circular profile from disorder-induced radial localization. An infinite-half-line Lyapunov exponent supplies the primary radial diagnostic, while Dirichlet, Neumann, and Robin calculations test the boundary insensitivity of selected bulk states. Transfer matrices and finite-ring scaling establish the angular localization length and operational crossover coefficient. Finally, a direct polar-grid calculation tracks the $\pm m_0$ subspace continuously from the symmetric limit and compares it with the projected ring at matched angular energy. This last calculation averages over independent radial as well as angular disorder, removing the state-selection and single-radial-realization biases that would otherwise obscure the projection test.

The article is organized as follows. Section~\ref{sec:model} introduces the continuum model and notation. Section~\ref{sec:radial} establishes disorder-induced radial localization and its boundary independence. Sections~\ref{sec:angular} and \ref{sec:curvature} derive and test the angular crossover law. Section~\ref{sec:notransition} presents diagnostics that prevent a false transition claim. Section~\ref{sec:comparison} compares the mechanism with tree and hyperbolic-lattice localization. Technical derivations and numerical details are collected in the appendices.

\section{Continuum model and notation}
\label{sec:model}

\subsection{Constant-curvature geometry}

\begin{figure}[t]
 \includegraphics[width=0.8\columnwidth]{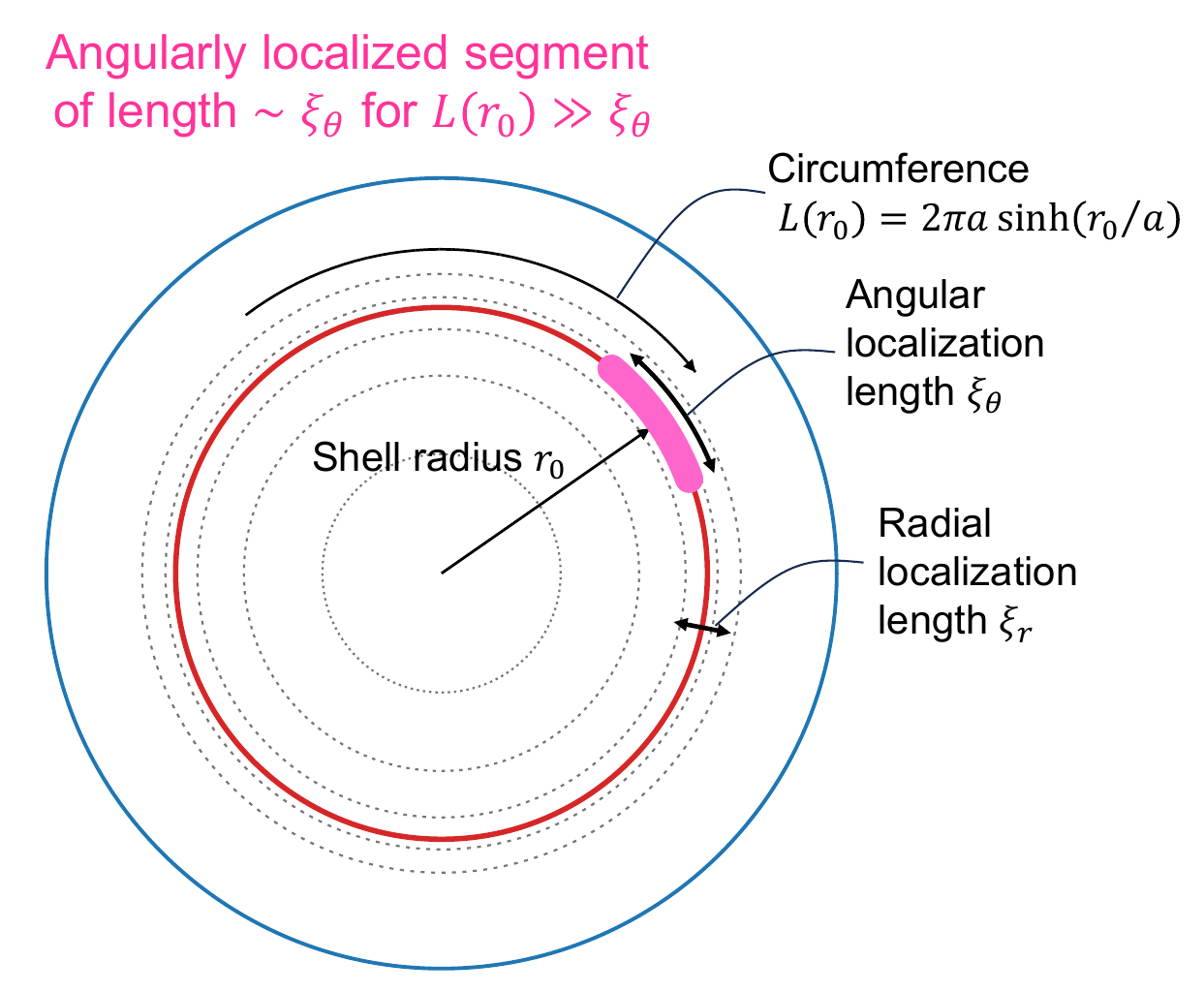}
 \caption{Geometry and localization mechanisms. A random radial potential first produces a mode whose envelope is localized around a geodesic radius $r_0$ with radial localization length $\xi_r$. Exact rotational symmetry makes this mode extended around the associated circle. A weak angle-dependent perturbation then localizes the projected mode along the physical arc coordinate $s$, with angular localization length $\xi_\theta$. The Poincar\'e-disk edge is the conformal boundary at infinite geodesic distance, not a finite outer wall.}
 \label{fig:geometry}
\end{figure}

We consider a rotationally symmetric two-dimensional Riemannian surface with constant Gaussian curvature $K$. In geodesic polar coordinates, $r\geq0$ denotes geodesic distance from a chosen origin and $\theta\in[0,2\pi)$ denotes the polar angle. The line element is
\begin{equation}
 \dd s^2=\dd r^2+S_K^2(r)\dd\theta^2,
 \label{eq:metric}
\end{equation}
where the metric radius $S_K(r)$ is
\begin{equation}
S_K(r)=
\begin{cases}
 a\sinh(r/a), & K=-a^{-2},\\
 r, & K=0,\\
 a\sin(r/a), & K=+a^{-2}.
\end{cases}
\label{eq:SK}
\end{equation}
Here $a>0$ is the curvature radius for nonzero $K$. The area element is $\dd A=S_K(r)\dd r\dd\theta$, the circumference of the geodesic circle at radius $r$ is
\begin{equation}
 L_K(r)=2\pi S_K(r),
 \label{eq:circumference}
\end{equation}
and the Laplace--Beltrami operator is
\begin{equation}
 \Delta_K=\partial_r^2+\frac{S_K'(r)}{S_K(r)}\partial_r
 +\frac{1}{S_K^2(r)}\partial_\theta^2.
 \label{eq:LB}
\end{equation}
A prime denotes differentiation with respect to $r$. Figure~\ref{fig:geometry} illustrates the negatively curved case.

\subsection{Random Schr\"odinger operator}

The single-particle Hamiltonian is
\begin{equation}
 H=-\frac{\hbar^2}{2M}\Delta_K+V_r(r)+\varepsilon V_\perp(r,\theta),
 \label{eq:Hfull}
\end{equation}
where $M$ is the particle mass, $\hbar$ is the reduced Planck constant, $V_r(r)$ is the dominant radial random potential, $V_\perp(r,\theta)$ is an angle-dependent random perturbation, and the dimensionless amplitude $\varepsilon\geq0$ controls rotational-symmetry breaking. The stationary wave function $\psi(r,\theta)$ satisfies $H\psi=E\psi$, where $E$ is the energy.

The radial disorder has zero mean and a finite radial correlation length $\ell_r$. The angle-dependent ensemble is stationary in physical arc length rather than in the coordinate angle. A local constant-curvature covariance that makes this assumption explicit is
\begin{equation}
\begin{split}
&\avg{V_\perp(r,\theta)V_\perp(r',\theta')}\\
&\quad=C_r^\perp(r-r')\,
C_\theta\!\left[S_K(\bar r)d_{2\pi}(\theta-\theta')\right],
\qquad \bar r=\frac{r+r'}{2},
\end{split}
\label{eq:Vperpcovariance}
\end{equation}
where $d_{2\pi}(\varphi)=\min_{n\in\mathbb Z}|\varphi+2\pi n|$. Thus the physical arc correlation length $\ell_\theta$ is independent of shell radius. The projected angular disorder introduced below has zero mean and correlation function $C_\theta(s)$. Its one-dimensional spectral density is
\begin{equation}
 \widetilde C_\theta(q)=\int_{-\infty}^{\infty}C_\theta(s)\e^{-\ii q s}\dd s,
 \label{eq:angularspectrum}
\end{equation}
where $q$ is the arc wave number and $\ii^2=-1$. If an implementation instead holds an angular correlation width fixed in $\theta$, then $\xi_\theta$ becomes radius dependent and the crossover criterion must be written self-consistently as $L_K(r_*)=c_Q\xi_\theta(r_*)$. Unless a dimensionful expression is required, the numerical calculations use $\hbar^2/(2M)=1$ and unit discretization spacing.

\subsection{Exact radial reduction}

At $\varepsilon=0$, continuous rotational symmetry permits the expansion
\begin{equation}
 \psi(r,\theta)=\sum_{m\in\mathbb Z}
 \frac{\e^{\ii m\theta}}{\sqrt{2\pi S_K(r)}}u_m(r),
 \label{eq:partialwave}
\end{equation}
where the integer $m$ is the angular-momentum quantum number and $u_m(r)$ is the corresponding one-dimensional radial amplitude. The prefactor gives the norm identity
\begin{equation}
 \int |\psi|^2\dd A=\sum_m\int_0^\infty |u_m(r)|^2\dd r.
 \label{eq:norm}
\end{equation}
Each channel satisfies
\begin{equation}
 \left[-\frac{\hbar^2}{2M}\frac{\dd^2}{\dd r^2}
 +V_r(r)+U_{m,K}(r)\right]u_m(r)=E u_m(r),
 \label{eq:radialgeneral}
\end{equation}
with geometric potential
\begin{equation}
 U_{m,K}(r)=\frac{\hbar^2}{2M}
 \left[\frac{m^2}{S_K^2(r)}+\frac{S_K''(r)}{2S_K(r)}
 -\frac{[S_K'(r)]^2}{4S_K^2(r)}\right].
 \label{eq:geometricpotential}
\end{equation}
For $K=-a^{-2}$,
\begin{equation}
 U_{m,-}(r)=\frac{\hbar^2}{2Ma^2}
 \left[\frac14+\frac{m^2-\frac14}{\sinh^2(r/a)}\right].
 \label{eq:Uhyper}
\end{equation}
Thus, at $r\gg a$, every fixed-$m$ channel approaches the same one-dimensional random Hamiltonian, shifted by the curvature energy $\hbar^2/(8Ma^2)$.

The angularly uniform shell profile follows directly from Eq.~\eqref{eq:partialwave}; it is a symmetry statement. Radial Anderson localization is separately characterized by the positive Lyapunov exponent $\gamma_r(E,m)$ of Eq.~\eqref{eq:radialgeneral}. The radial localization length is
\begin{equation}
 \xi_r(E,m)=\gamma_r^{-1}(E,m),
 \label{eq:xir}
\end{equation}
and a localized envelope centered at $r_0$ behaves asymptotically as $|u_m(r)|\propto\exp[-|r-r_0|/\xi_r]$. This distinction prevents the shell shape from being mistaken for the localization mechanism itself.

\section{Radial Anderson localization and boundary independence}
\label{sec:radial}

\subsection{Infinite-half-line Lyapunov exponent and controls}

In the asymptotic hyperbolic region we test the reduced equation in units $\hbar^2/(2M)=1$,
\begin{equation}
 -u''(r)+V_r(r)u(r)=k_r^2u(r),
 \label{eq:radialtest}
\end{equation}
where $k_r>0$ is the radial wave number and a double prime denotes the second derivative with respect to $r$. The potential is piecewise constant in cells of length $\ell_r$ and independently uniform in $[-W_r/2,W_r/2]$, where $W_r$ is the radial disorder width. Randomizing the cell origin gives
\begin{equation}
 \widetilde C_r(q)=\frac{W_r^2\ell_r}{12}
 \sinc^2\!\left(\frac{q\ell_r}{2}\right),
 \qquad \sinc x=\frac{\sin x}{x},
 \label{eq:Cr}
\end{equation}
where $\widetilde C_r(q)$ is the radial disorder spectrum. Weak-disorder perturbation theory gives~\cite{Lifshits1988,IzrailevKrokhin1999}
\begin{equation}
 \gamma_r^{\rm B}(k_r)=\frac{\widetilde C_r(2k_r)}{8k_r^2}
 =\frac{W_r^2\ell_r}{96k_r^2}\sinc^2(k_r\ell_r),
 \label{eq:bornradial}
\end{equation}
where the superscript ${\rm B}$ denotes the Born approximation.

\begin{figure*}[t]
 \includegraphics[width=0.94\textwidth]{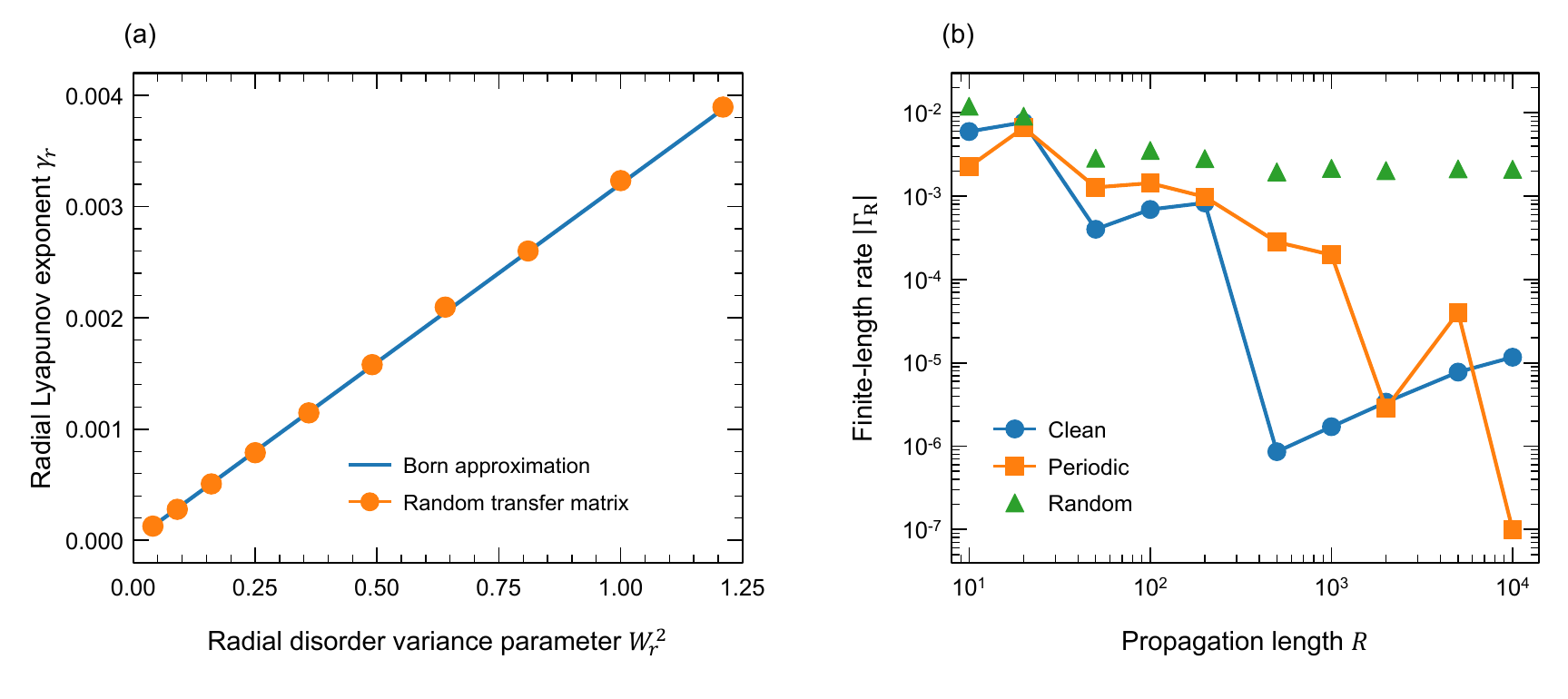}
 \caption{Disorder-induced radial localization. (a) Radial Lyapunov exponent $\gamma_r$ versus squared disorder width $W_r^2$. Symbols are infinite-half-line transfer-matrix results; the line is the Born approximation in Eq.~\eqref{eq:bornradial}. (b) Absolute finite-length rate $|\Gamma_R|$ for clean, periodic, and random radial potentials. The first two approach zero, while the random ensemble approaches a positive asymptotic rate. Error bars show standard errors where disorder averaging is used.}
 \label{fig:radialcontrols}
\end{figure*}

The numerical Lyapunov exponent is calculated from the product $\mathcal T_R$ of exact cell-transfer matrices over propagation length $R$,
\begin{equation}
 \gamma_r=\lim_{R\rightarrow\infty}\frac{1}{R}\avg{\ln\Vert\mathcal T_R\Vert}.
 \label{eq:lyapunovdefinition}
\end{equation}
The brackets denote an ensemble average, and $\Vert\mathcal T_R\Vert$ denotes the matrix spectral norm, i.e., the largest singular value of the transfer matrix $\mathcal T_R$. This is the norm used in all transfer-matrix Lyapunov exponents below. Figure~\ref{fig:radialcontrols}(a) shows agreement between Eq.~\eqref{eq:bornradial} and 120-realization transfer-matrix data for $k_r=1.2$, $\ell_r=0.5$, and $R=2.0\times10^4$. The fitted quadratic coefficient agrees with the Born value throughout the displayed weak-disorder range.

To separate disorder localization from a circular profile imposed by symmetry, Fig.~\ref{fig:radialcontrols}(b) compares three radial ensembles at the same modulation scale: a clean potential, a period-two potential, and independent random cells. We plot the finite-length rate
\begin{equation}
 \Gamma_R=R^{-1}\ln\Vert\mathcal T_R\Vert.
 \label{eq:finitegamma}
\end{equation}
The clean and periodic rates decay toward zero within their propagating bands, whereas the random rate approaches a positive constant. This control shows that axisymmetry determines only the angular shape; radial exponential localization requires random scattering.

\subsection{Finite-disk boundary test}
\label{subsec:boundary}

A finite hyperbolic disk has a nonvanishing boundary-to-area ratio, so global open-disk spectral statistics can be dominated by edge states. Our primary result, Eq.~\eqref{eq:lyapunovdefinition}, avoids this problem because it is defined on an infinite half-line. We nevertheless perform a separate boundary-sensitivity test for a localized radial eigenstate.

The radial finite-difference Hamiltonian has unit spacing and a fixed Dirichlet condition at the inner endpoint. At the outer endpoint $r=R$, we impose in turn: Dirichlet, $u(R)=0$; Neumann, $u'(R)=0$; and Robin, $u'(R)+\kappa_bu(R)=0$, where $\kappa_b=0.5$ in lattice units. For each disorder realization, a reference state is selected under the Dirichlet condition near a fixed target energy. Its center is
\begin{equation}
 \bar r=\sum_j r_j|u_j|^2,
 \label{eq:center}
\end{equation}
where $j$ labels radial grid points and $u_j$ is the normalized lattice amplitude. Its operational inverse-participation localization length is
\begin{equation}
 \xi_{\rm IPR}=\left(\sum_j|u_j|^4\right)^{-1},
 \label{eq:xiipr}
\end{equation}
and its distance to the outer boundary is $d_b=R-\bar r$. The state is matched across boundary conditions by maximum wave-function overlap in the bulk core.

\begin{figure*}[t]
 \includegraphics[width=0.94\textwidth]{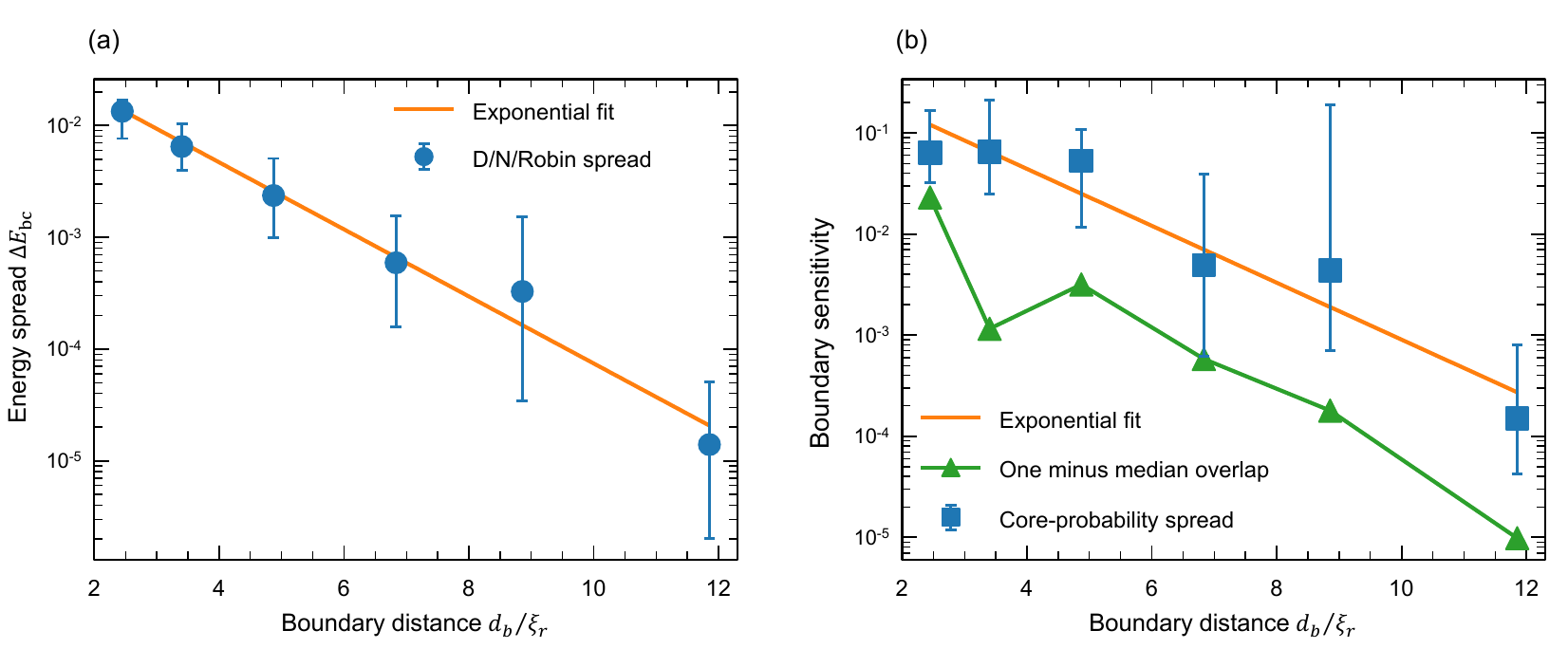}
 \caption{Boundary-condition robustness of radially localized bulk states. (a) Median three-boundary energy spread $\Delta E_{\rm bc}$ versus the boundary distance $d_b$ measured in units of the inverse-participation length $\xi_{\rm IPR}$. (b) Median core-density spread and matched-state overlap. Symbols summarize independent disorder realizations; bands or error bars indicate interquartile ranges. Straight lines in semilogarithmic coordinates are the fits in Eqs.~\eqref{eq:boundaryenergyfit} and \eqref{eq:boundarycorefit}.}
 \label{fig:boundary}
\end{figure*}

Figure~\ref{fig:boundary}(a) shows the median energy spread among the three boundary conditions versus the median ratio $d_b/\xi_{\rm IPR}$. Figure~\ref{fig:boundary}(b) shows the analogous spread of the core probability density and the median overlap. The fitted laws are
\begin{align}
 \ln\Delta E_{\rm bc}&=-2.60-0.691\,d_b/\xi_{\rm IPR},
 \label{eq:boundaryenergyfit}\\
 \ln\Delta P_{\rm core}&=-0.530-0.648\,d_b/\xi_{\rm IPR},
 \label{eq:boundarycorefit}
\end{align}
where $\Delta E_{\rm bc}$ is the maximum energy difference among the three matched states and $\Delta P_{\rm core}$ is the corresponding core-density spread. At $d_b/\xi_{\rm IPR}\simeq11.9$, the median energy spread is $1.4\times10^{-5}$ and the median overlap is $0.99999$. Hence states satisfying a conservative bulk criterion $d_b\gtrsim7\xi_{\rm IPR}$ are insensitive to the outer boundary within the present resolution. The numerical exponent is operational because $\xi_{\rm IPR}$ is not the asymptotic tail length; the essential result is the exponential suppression under all three boundary conditions.

\section{Angular localization of a projected shell}
\label{sec:angular}

\subsection{Effective one-dimensional ring}

Let $u_\alpha(r)$ be a normalized radial localized mode labeled by $\alpha$, centered at $r_0$ and having radial localization length $\xi_r$. We assume a narrow shell,
\begin{equation}
 \xi_r\ll a,\qquad \xi_r\ll S_K(r_0),
 \label{eq:narrowshell}
\end{equation}
and weak inter-radial-mode mixing. Let
\begin{equation}
 P_\alpha=|u_\alpha\rangle\langle u_\alpha|\otimes I_\theta,
 \qquad Q_\alpha=I-P_\alpha .
 \label{eq:projectors}
\end{equation}
The diagonal projection of Eq.~\eqref{eq:Hfull} is
\begin{equation}
 P_\alpha H P_\alpha=E_\alpha^{(r)}-
 \frac{\hbar^2\beta_\alpha}{2M}\partial_\theta^2
 +\varepsilon V_\alpha(\theta),
 \label{eq:projectedtheta}
\end{equation}
where $E_\alpha^{(r)}$ is the expectation value of the radial part,
\begin{equation}
 \beta_\alpha=\int_0^\infty
 \frac{|u_\alpha(r)|^2}{S_K^2(r)}\dd r,
 \qquad S_\alpha=\beta_\alpha^{-1/2},
 \label{eq:betaalpha}
\end{equation}
\begin{equation}
 V_\alpha(\theta)=\int_0^\infty |u_\alpha(r)|^2V_\perp(r,\theta)\dd r
 \label{eq:Valpha}
\end{equation}
is the projected angular random potential. Coupling omitted from the closed ring Hamiltonian is instead
\begin{equation}
 H_{\rm mix}=Q_\alpha H P_\alpha+P_\alpha H Q_\alpha.
 \label{eq:Hmix}
\end{equation}
For a radial spectral separation $\Delta_\alpha$ in the energy window of interest, a useful control parameter is
\begin{equation}
 \mu_\alpha=\frac{\lVert Q_\alpha H P_\alpha\rVert}{\Delta_\alpha}\ll1.
 \label{eq:mixingparameter}
\end{equation}
Equation~\eqref{eq:betaalpha}, rather than the point value $S_K^{-2}(r_0)$, is the exact diagonal projection of the angular kinetic energy. The narrow-shell conditions imply $S_\alpha=S_K(r_0)[1+O(\delta_{\rm met})]$, where a useful dimensionless measure of metric variation is
\begin{equation}
 \delta_{\rm met}=
 \left\{\int_0^\infty |u_\alpha(r)|^2
 \left[\frac{S_\alpha^2}{S_K^2(r)}-1\right]^2\dd r\right\}^{1/2}.
 \label{eq:deltametric}
\end{equation}

The effective arc coordinate is
\begin{equation}
 s=S_\alpha\theta,\qquad 0\leq s<2\pi S_\alpha.
 \label{eq:arc}
\end{equation}
In this coordinate, Eq.~\eqref{eq:projectedtheta} becomes the standard one-dimensional random-ring Hamiltonian
\begin{equation}
 H_\alpha=E_\alpha^{(r)}-\frac{\hbar^2}{2M}\partial_s^2+
 \varepsilon V_\alpha(s).
 \label{eq:ringH}
\end{equation}
The physical shell circumference $L_K(r_0)=2\pi S_K(r_0)$ and the projected circumference $2\pi S_\alpha$ therefore agree up to the explicitly controlled metric error. The conditions in Eqs.~\eqref{eq:narrowshell} and \eqref{eq:mixingparameter} have distinct roles: the former controls variation of the metric across the shell, whereas the latter controls coupling to other radial modes. Both are tested directly below.
For an arc wave number $k_\theta>0$, the weak-disorder angular Lyapunov exponent is
\begin{equation}
 \gamma_\theta(k_\theta)=\xi_\theta^{-1}
 =\frac{\varepsilon^2\widetilde C_\theta(2k_\theta)}{8k_\theta^2}
 +O(\varepsilon^3),
 \label{eq:borntheta}
\end{equation}
where $\xi_\theta$ is the angular localization length in physical arc units. Thus $\xi_\theta\propto\varepsilon^{-2}$ at weak perturbation. For the sign-symmetric uniform distribution used numerically, odd disorder moments vanish and the remainder starts at $O(\varepsilon^4)$.

We quantify angular extension on a ring of circumference $L$ by the lattice participation length
\begin{equation}
 P_\theta=\left(\sum_{j=1}^{N}|\phi_j|^4\right)^{-1},
 \label{eq:Ptheta}
\end{equation}
where $N=L$ in the unit-spacing discretization and $\phi_j$ is a normalized projected eigenvector. For a clean standing wave in the chosen energy window, $P_\theta/N\simeq2/3$. We therefore define the normalized shell-extension indicator
\begin{equation}
 Q_{\rm sh}=\frac{P_\theta/N}{2/3}.
 \label{eq:Qshell}
\end{equation}
A rotationally extended shell has $Q_{\rm sh}\simeq1$, while $Q_{\rm sh}$ decreases when the circumference exceeds the angular localization length.

\subsection{Transfer matrix and one-parameter scaling}

The numerical ring model is
\begin{equation}
 -\phi_{j+1}-\phi_{j-1}+[2+\varepsilon w_j]\phi_j=E_\theta\phi_j,
 \label{eq:ringlattice}
\end{equation}
with periodic indexing, independent $w_j\in[-1/2,1/2]$, and target energy $E_\theta=0.5$. The unit-spacing Born coefficient is $A_{\rm B}=1/42=0.0238095$ for
\begin{equation}
 \gamma_\theta=A\varepsilon^2.
 \label{eq:Agamma}
\end{equation}
Transfer matrices over $1.2\times10^5$ sites and 120 realizations give
\begin{equation}
 A=0.0241198\pm0.0000272,
 \label{eq:Afit}
\end{equation}
which differs from $A_{\rm B}$ by about $1.3\%$. The uncertainty in Eq.~\eqref{eq:Afit} is the conditional regression error over the finite perturbation window, not an uncertainty on the $\varepsilon\rightarrow0$ intercept. The longer weak-disorder calculation reported in Fig.~\ref{fig:weaklog} below separates that intercept from finite-$\varepsilon$ corrections and recovers $A_{\rm B}$.

\begin{figure*}[t]
 \includegraphics[width=0.94\textwidth]{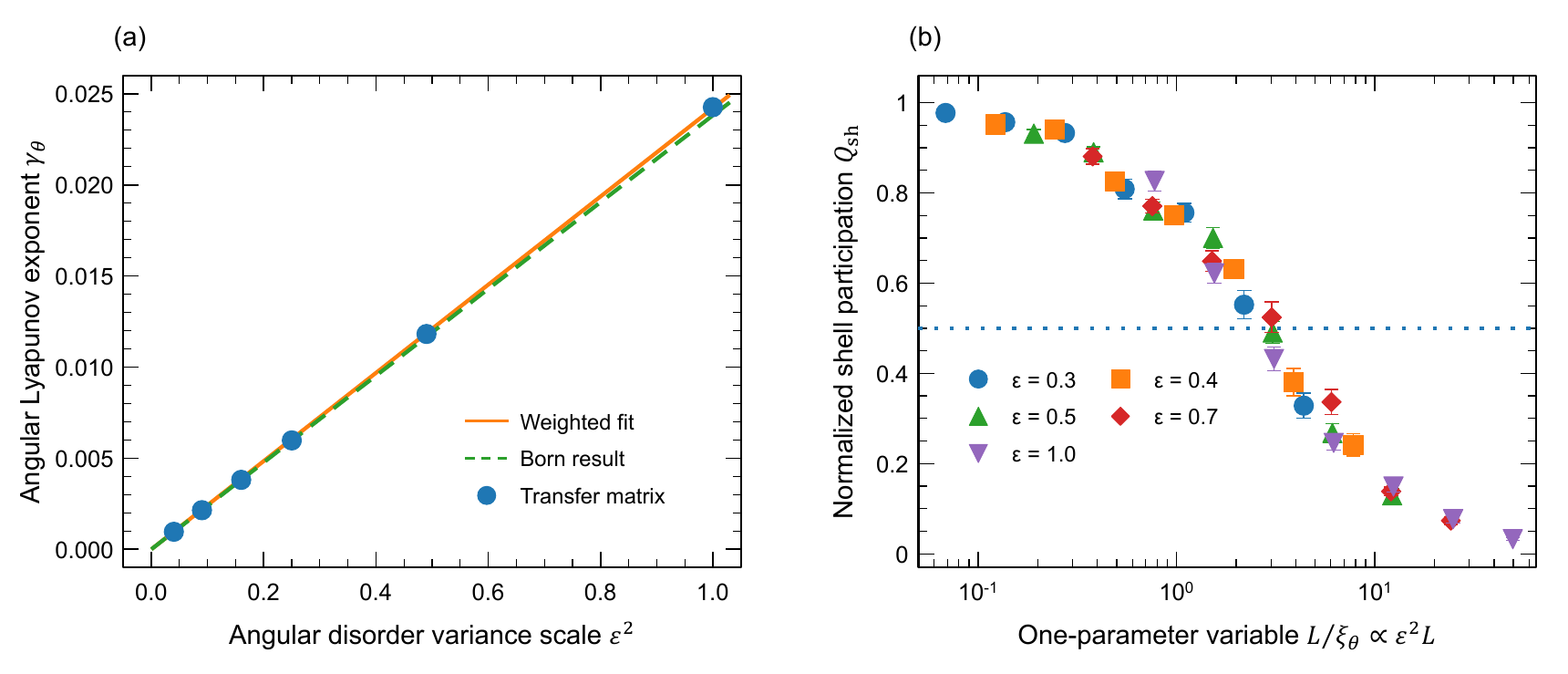}
 \caption{Angular localization of the projected shell. (a) Angular Lyapunov exponent $\gamma_\theta$ versus $\varepsilon^2$. The displayed finite-window quadratic coefficient is $1.3\%$ above the discrete Born value; its weak-disorder intercept is resolved independently in Fig.~\ref{fig:weaklog}. (b) Normalized shell-extension indicator $Q_{\rm sh}$ versus the one-parameter scaling variable $L/\xi_\theta$. In the unit-spacing discretization $L=N$, so this is the same variable written as $N/\xi_\theta$ in the text. The common collapse demonstrates a circumference-controlled crossover.}
 \label{fig:angular}
\end{figure*}

Figure~\ref{fig:angular}(a) displays the quadratic law. Figure~\ref{fig:angular}(b) collapses the participation data for five perturbation strengths when plotted against $N/\xi_\theta$. The collapse shows that circumference, rather than disk radius or an outer boundary, controls the shell breakdown. Defining an operational crossover by $Q_{\rm sh}=1/2$ gives
\begin{equation}
 N_*=c_Q\xi_\theta,
 \qquad c_Q=2.81\pm0.34,
 \label{eq:cQ}
\end{equation}
where $N_*$ is the crossover circumference and the uncertainty is the standard deviation across the five perturbation strengths. The coefficient $c_Q$ depends on the chosen threshold and observable; the curvature dependence derived below does not.

\section{Constant-curvature crossover law}
\label{sec:curvature}

The operational criterion in Eq.~\eqref{eq:cQ} becomes
\begin{equation}
 L_K(r_*)=c_Q\xi_\theta,
 \label{eq:criterion}
\end{equation}
where $r_*$ is the geodesic radius at which a shell crosses the chosen angular-localization threshold. Substitution of Eqs.~\eqref{eq:SK} and \eqref{eq:circumference} gives
\begin{align}
 K=-a^{-2}:\quad
 r_*&=a\arsinh\!\left(\frac{c_Q\xi_\theta}{2\pi a}\right),
 \label{eq:rhyper}\\
 K=0:\quad
 r_*&=\frac{c_Q\xi_\theta}{2\pi},
 \label{eq:rflat}\\
 K=+a^{-2}:\quad
 r_*&=a\arcsin\!\left(\frac{c_Q\xi_\theta}{2\pi a}\right),
 \label{eq:rpositive}
\end{align}
where Eq.~\eqref{eq:rpositive} exists only when $c_Q\xi_\theta\leq2\pi a$ within a hemisphere. Positive curvature can therefore prevent the threshold from being reached for sufficiently weak angular disorder.

For negative curvature, define the dimensionless crossover variable
\begin{equation}
 z=\frac{c_Q\xi_\theta}{2\pi a}.
 \label{eq:z}
\end{equation}
Equation~\eqref{eq:rhyper} becomes $r_*/a=\arsinh z$. Its two limits are
\begin{align}
 z\ll1:&\quad r_*\simeq\frac{c_Q\xi_\theta}{2\pi},
 \label{eq:localflat}\\
 z\gg1:&\quad r_*\simeq a\ln(2z).
 \label{eq:hyperasym}
\end{align}
The first expression is the locally flat regime. Using $\xi_\theta=(A\varepsilon^2)^{-1}$ in the second gives
\begin{equation}
 r_*=2a\ln\frac{1}{\varepsilon}
 +a\ln\!\left(\frac{c_Q}{\pi a A}\right)+O(\varepsilon^2).
 \label{eq:loglaw}
\end{equation}
In contrast, Eq.~\eqref{eq:rflat} gives $r_*\propto\varepsilon^{-2}$ in the Euclidean plane. The continuum-specific result is therefore not exponential volume growth alone, which also occurs on trees, but the exact interpolation from local two-dimensional flat geometry to a logarithmic large-distance law through the smooth metric factor $S_K(r)$. The singular character of this crossover can be stated as the noncommutation of the weak-disorder and flat-curvature limits:
\begin{equation}
\begin{split}
 \lim_{a\rightarrow\infty}\lim_{\varepsilon\rightarrow0}
 \frac{r_*}{\xi_\theta}&=0,\\
 \lim_{\varepsilon\rightarrow0}\lim_{a\rightarrow\infty}
 \frac{r_*}{\xi_\theta}&=\frac{c_Q}{2\pi}.
\end{split}
\label{eq:noncommutinglimits}
\end{equation}
For fixed curvature radius, the logarithm grows more slowly than $\xi_\theta$; taking the flat limit first instead recovers the Euclidean proportionality.

\begin{figure*}[t]
 \includegraphics[width=0.94\textwidth]{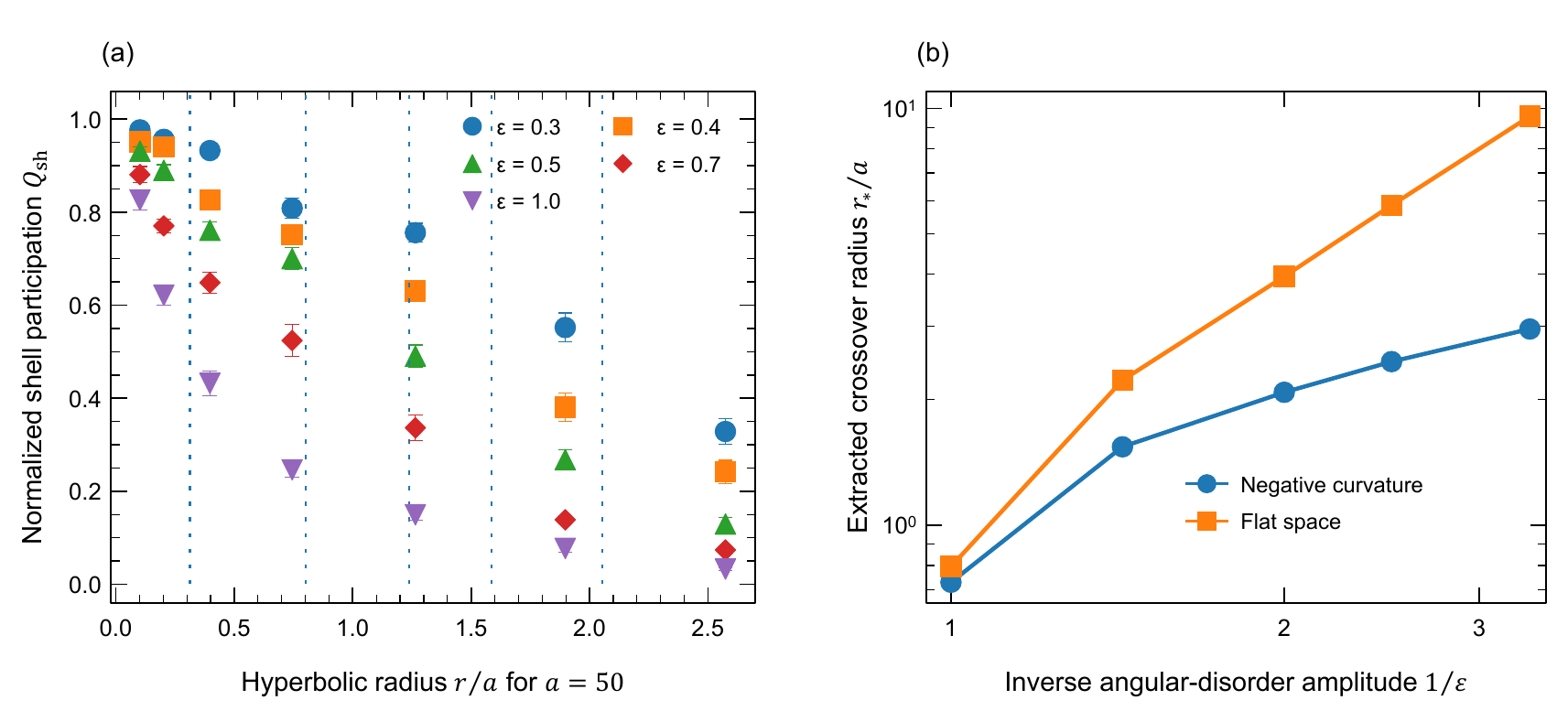}
 \caption{Extraction and geometric conversion of the shell crossover. (a) The vertical markers identify the hyperbolic crossover radii $r_*$ obtained by converting the extracted circumferences $N_*$ at which $Q_{\rm sh}=1/2$. (b) The same $N_*$ values converted to $r_*$ in flat and hyperbolic geometry. The flat conversion grows algebraically as the perturbation weakens, while the hyperbolic conversion bends to the logarithmic law in Eq.~\eqref{eq:loglaw}.}
 \label{fig:extraction}
\end{figure*}

\begin{figure}[t]
 \includegraphics[width=\columnwidth]{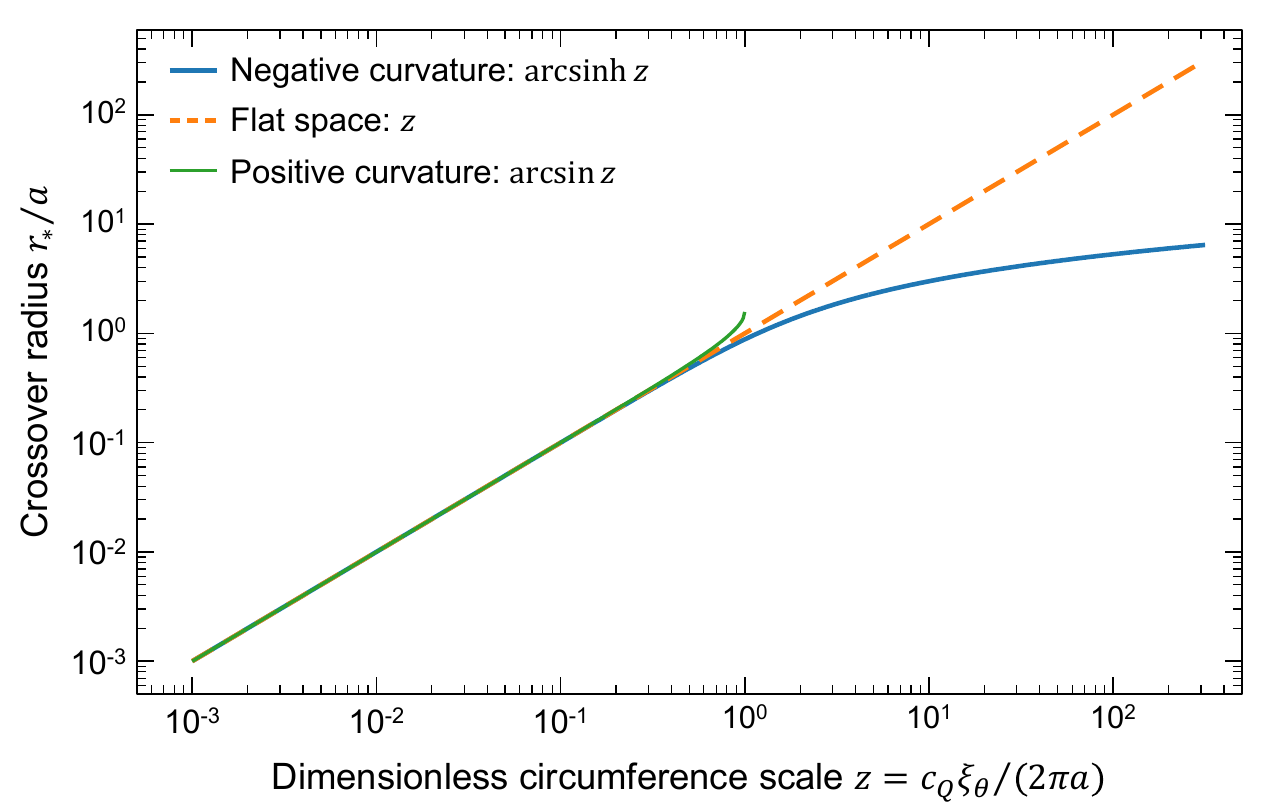}
 \caption{Unified constant-curvature conversion obtained from Eq.~\eqref{eq:criterion}. The dimensionless crossover radius $r_*/a$ is plotted against $z=c_Q\xi_\theta/(2\pi a)$. Negative curvature gives $\arsinh z$ and hence a logarithmic large-$z$ regime; flat space gives $z$; and positive curvature gives $\arcsin z$, which exists only for $z\leq1$ within a hemisphere.}
 \label{fig:curvature}
\end{figure}

Figure~\ref{fig:extraction} maps the numerically extracted ring crossovers into Euclidean and hyperbolic radii. Figure~\ref{fig:curvature} shows the full constant-curvature conversion. The same ring data generate all curves; only the continuum metric changes. This separates the universal one-dimensional angular localization problem from its geometric realization.

\section{Why the result is a crossover rather than a transition}
\label{sec:notransition}

Angular-momentum mixing and radial delocalization are distinct phenomena. For the ring-correlated ensemble considered here, the loss of rotational extension is described by a projected one-dimensional localization problem and should not be reinterpreted as a metal--insulator transition. This statement is deliberately restricted to the present ensemble and crossover observable; the thermodynamic transport problem for generic isotropic disorder on the hyperbolic plane lies outside the scope of this work.

\subsection{Typical local density of states}

\begin{figure}[!b]
 \includegraphics[width=\columnwidth]{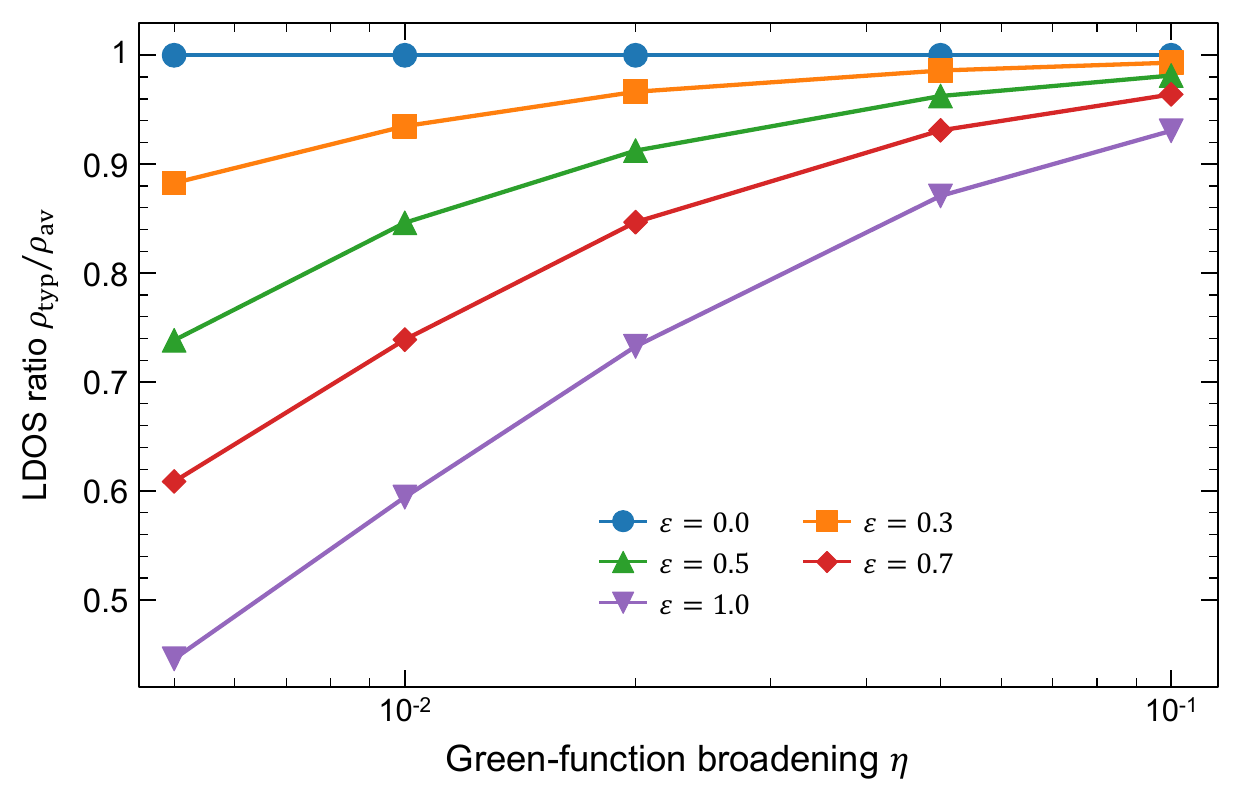}
 \caption{Illustrative ratio of typical to average local density of states, $\rho_{\rm typ}/\rho_{\rm av}$, versus spectral broadening $\eta$ for the projected periodic chain at $E_\theta=0.5$. The clean ratio remains unity, while nonzero angular disorder produces a continuous decrease with decreasing $\eta$. Because the chain length is fixed, this panel is not a finite-size transition test.}
 \label{fig:ldos}
\end{figure}

For the angular lattice Hamiltonian, the local density of states at site $j$ is
\begin{equation}
 \rho_j(E,\eta)=-\frac{1}{\pi}\operatorname{Im}G_{jj}(E+\ii\eta),
 \label{eq:ldos}
\end{equation}
where $G(z)=(z-H)^{-1}$ is the resolvent and $\eta>0$ is a spectral broadening. The geometric mean is widely used as a typical local spectral measure in localization problems~\cite{Dobrosavljevic2003}. The arithmetic and geometric means are
\begin{align}
 \rho_{\rm av}(E,\eta)&=\avg{\rho_j(E,\eta)},
 \label{eq:rhoav}\\
 \rho_{\rm typ}(E,\eta)&=\exp\avg{\ln\rho_j(E,\eta)}.
 \label{eq:rhotyp}
\end{align}
Figure~\ref{fig:ldos} plots $\rho_{\rm typ}/\rho_{\rm av}$ at $E_\theta=0.5$ for periodic chains of 4096 sites and 48 disorder realizations. The clean ratio is unity. For nonzero $\varepsilon$, the ratio decreases continuously as $\eta$ is reduced, as expected for a one-dimensional localized spectrum. This fixed-size broadening scan is illustrative and is not used to infer the absence of a transition: the positive one-dimensional Lyapunov exponent for every nonzero short-range disorder amplitude already establishes localization in the projected model. A transition claim for generic isotropic hyperbolic disorder would instead require a distinct multi-size transport and order-of-limits analysis.

\subsection{Direct polar-grid test of the projection}

We next compare the projected ring with a direct discretization of the same two-dimensional polar Hamiltonian. Writing $\chi=S_K^{1/2}\psi$ and using units $\hbar^2/(2M)=1$, the hyperbolic operator is
\begin{equation}
 H_\chi=-\partial_r^2-\frac{1}{S_K^2(r)}\partial_\theta^2
 +U_g(r)+V_r(r)+\varepsilon V_\perp(r,\theta),
 \label{eq:fullpolargrid}
\end{equation}
where
\begin{equation}
 U_g(r)=\frac{1}{4a^2}
 -\frac{1}{4a^2\sinh^2(r/a)}.
 \label{eq:Ugpolar}
\end{equation}
Centered radial differences, periodic angular differences, and Dirichlet radial endpoints are used. The radial potential is real and cellwise uniform, common to every angle, while $V_\perp$ is real and independently uniform in $[-1/2,1/2]$ on every polar cell. The local physical angular cell size is approximately unity across the narrow selected shell. Crucially, the projected ring and the full matrix use the same $V_\perp(r,\theta)$ realization in every comparison.

The reference calculation uses $a=100$, radial interval $5<r<37$, radial spacing $h=0.5$, unit-width radial-disorder cells with disorder width $W_r=16$, $N_\theta=128$, and angular harmonic $m_0=15$. Six radial-disorder realizations are fixed before any angular perturbation is generated. In each realization, the admissible bulk mode whose discrete $m_0$ angular energy is closest to the ring target $E_\theta=0.5$ is selected at $\varepsilon=0$. Across the six modes,
\begin{equation}
 \begin{gathered}
 \avg{r_0}=20.605,\\
 \avg{\xi_{\rm IPR}}=1.776,\\
 0.0166\leq\xi_{\rm IPR}/a\leq0.0188,\\
 0.0801\leq\xi_{\rm IPR}/S_K(r_0)\leq0.0899,\\
 8.37\leq d_{\min}/\xi_{\rm IPR}\leq9.38,
 \end{gathered}
 \label{eq:projectioncriteria}
\end{equation}
where
\begin{equation}
 d_{\min}=\min(r_0-r_{\rm in},r_{\rm out}-r_0)
 \label{eq:dmin}
\end{equation}
tests both radial boundaries. The mean discrete angular energy is $0.50024$ (range $0.49560$--$0.50486$), matching the $E_\theta=0.5$ ring calculation within $0.9\%$ for every radial realization. The physical angular spacing is $1.013$--$1.022$, the rms metric-coefficient variation in Eq.~\eqref{eq:deltametric} is $0.0457$--$0.0507$, and the nearest radial spectral gap is $1.09$--$3.67$. These values resolve the envelope, keep it narrow relative to both curvature scales, and place it inside the conservative bulk criterion established in Fig.~\ref{fig:boundary}.

At $\varepsilon=0$, the two-dimensional subspace
\begin{equation}
 {\cal S}_0=\operatorname{span}\{
 u_\alpha(r)\e^{\ii m_0\theta},
 u_\alpha(r)\e^{-\ii m_0\theta}\}
 \label{eq:cleanSubspace}
\end{equation}
is fixed before observing any perturbed eigenstate. It is followed in steps of $\Delta\varepsilon=0.125$ by selecting the two-dimensional eigensubspace with maximum overlap with the subspace at the preceding step. The projected ring is tracked by the same rule. This continuity construction avoids selecting states by their final projection weight or participation.

For normalized full-grid eigenvectors $\psi_{nj}^{(\ell)}$ spanning the tracked subspace, define
\begin{equation}
 \phi_j^{(\ell)}=\sum_n u_{\alpha n}^{*}\psi_{nj}^{(\ell)},\qquad
 Z_{\rm sub}=\frac12\sum_{\ell=1}^{2}\sum_j
 |\phi_j^{(\ell)}|^2,
 \label{eq:projectionweight}
\end{equation}
where the discrete quadrature factor is absorbed into the normalized lattice vectors. The basis-independent angular density and extension measure are
\begin{equation}
 p_j^{\rm full}=\frac{1}{2Z_{\rm sub}}
 \sum_{\ell=1}^{2}|\phi_j^{(\ell)}|^2,\qquad
 Q_{\rm sub}^{\rm full}=
 \left[N_\theta\sum_j(p_j^{\rm full})^2\right]^{-1},
 \label{eq:Qsub}
\end{equation}
with the analogous definition for the projected ring. The clean $\pm m_0$ subspace has $Q_{\rm sub}=1$. After orthonormalizing the radially projected full subspace, its fidelity with the projected-ring subspace is
\begin{equation}
 F_{\rm sub}=\frac12
 \operatorname{Tr}(P_{\rm full}^{(\alpha)}P_{\rm proj}).
 \label{eq:Fsub}
\end{equation}

\begin{figure*}[t]
 \includegraphics[width=0.85\textwidth]{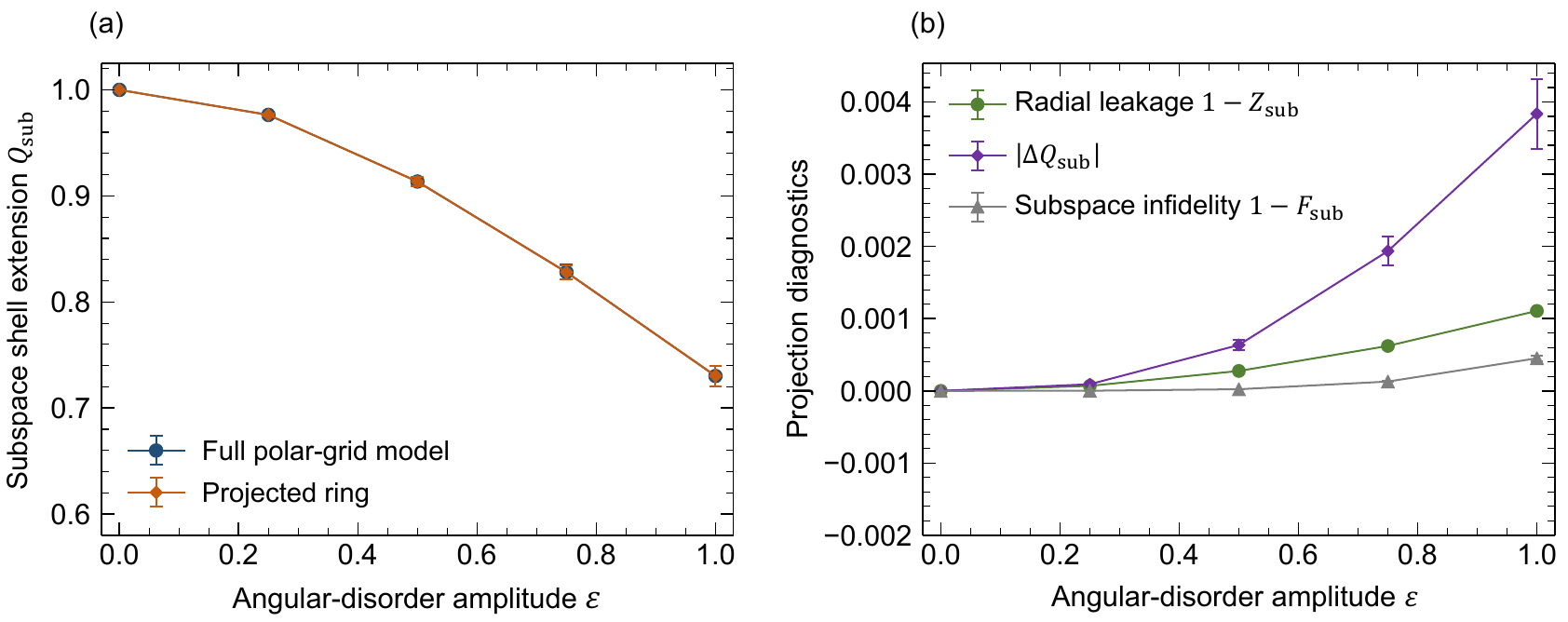}
 \caption{Projection validity at matched angular energy. (a) Basis-independent subspace shell extension from the full polar-grid model and projected ring using identical angular disorder. (b) Radial leakage, shell-extension mismatch, and subspace infidelity. Symbols are means over six independent radial-disorder realizations and eight angular-disorder realizations per radial sample (48 pairs); error bars are standard errors. State-resolved energies, tracking overlaps, random-seed records, and convergence series are available from the corresponding author upon reasonable request.}
 \label{fig:projection}
\end{figure*}

Figure~\ref{fig:projection} averages eight angular-disorder realizations for each of six independent radial-disorder realizations, giving 48 independent pairs at each displayed $\varepsilon$. At $\varepsilon=1$, the mean values are $\avg{1-Z_{\rm sub}}=1.11\times10^{-3}$, $\avg{1-F_{\rm sub}}=4.51\times10^{-4}$, and $\avg{|Q_{\rm sub}^{\rm full}-Q_{\rm sub}^{\rm proj}|}=3.84\times10^{-3}$.

Independent convergence checks use $h=0.625,0.500,0.400$, $N_\theta=112,128,144$, radial domains $4$--$38$, $5$--$37$, and $6$--$36$, and cumulative ensembles of $2\times2$, $4\times4$, and $6\times8$ radial-by-angular realizations. The same globally indexed radial cell disorder is retained in overlapping domains. At $\varepsilon=0.75$, the grid and domain variants give a maximum mean $Q_{\rm sub}$ mismatch of $3.0\times10^{-3}$ and maximum radial leakage of $9.8\times10^{-4}$. The test therefore validates the single-mode projection for the stated correlated ensemble and parameter window. It does not constitute a thermodynamic calculation for generic isotropic disorder.

\subsection{Weak-disorder extrapolation and logarithmic asymptote}

The finite-window coefficient in Eq.~\eqref{eq:Afit} is not used as a weak-limit test. We therefore repeat the transfer-matrix calculation at 12 perturbation strengths whose values of $\ln(1/\varepsilon)$ are equally spaced between $\varepsilon=0.9$ and $0.04$. Each point contains 192 independent realizations. The propagation length is chosen to give approximately 400 expected Lyapunov e-foldings, subject to a minimum of $5.0\times10^5$ steps; the longest run uses $1.05\times10^7$ steps.

For $\varepsilon\leq0.22$, a weighted fit to
\begin{equation}
 \frac{\gamma_\theta}{\varepsilon^2}
 =A_0+B\varepsilon^2
 \label{eq:weakfitform}
\end{equation}
gives
\begin{align}
 A_0&=0.0237813\pm0.0000447,\label{eq:A0weak}\\
 B&=0.00343\pm0.00191.\label{eq:Bweak}
\end{align}
Thus $A_0-A_{\rm B}=(-2.82\pm4.47)\times10^{-5}$: the extrapolated coefficient agrees with $1/42$ within one standard error. The weighted fit has $\chi^2/\mathrm{dof}=0.37$.

\begin{figure*}[t]
 \includegraphics[width=0.85\textwidth]{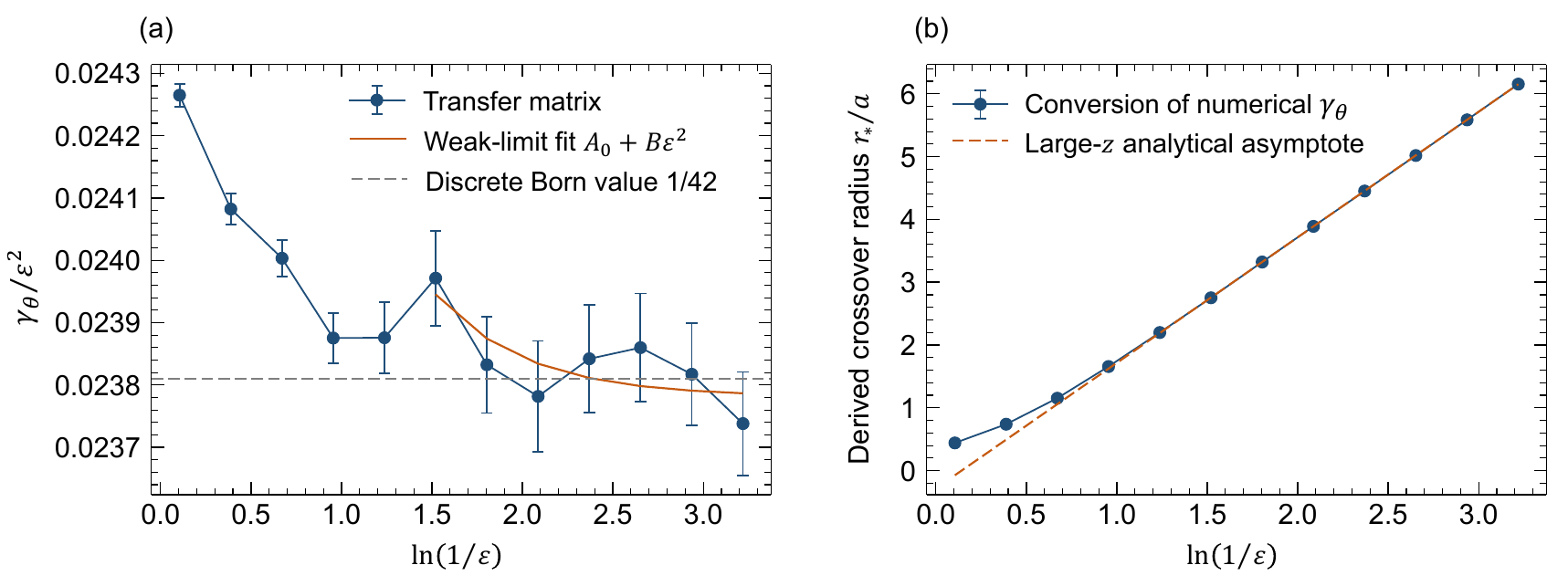}
 \caption{Weak-disorder coefficient and its geometric conversion. (a) $\gamma_\theta/\varepsilon^2$ versus $\ln(1/\varepsilon)$. The orange curve is the weighted fit $A_0+B\varepsilon^2$ over $\varepsilon\leq0.22$, and the gray line is the discrete Born coefficient $1/42$. (b) Equation~\eqref{eq:rhyper} applied to the same numerical $\gamma_\theta$ values, together with the large-$z$ analytical asymptote. This panel is a derived conversion, not an independent full-model test. The points use 192 realizations and up to $1.05\times10^7$ transfer steps; error bars are standard errors.}
 \label{fig:weaklog}
\end{figure*}

Figure~\ref{fig:weaklog}(b) converts the same numerical $\gamma_\theta$ values to $r_*/a$ using Eq.~\eqref{eq:rhyper}, $c_Q=2.81$, and $a=50$. As $\ln(1/\varepsilon)$ increases, this analytical conversion approaches Eq.~\eqref{eq:loglaw}. Panel (b) illustrates the geometric consequence of the transfer-matrix data; it is not an independent full-hyperbolic numerical calculation.

\section{Relation to Bethe and hyperbolic-lattice localization}
\label{sec:comparison}

Trees, regular hyperbolic lattices, and the continuous hyperbolic plane all have exponential growth, so resonance proliferation is not unique to the present system. Anderson localization on trees is commonly analyzed by Green-function recursion and competing exponential decay and branching~\cite{AbouChacra1973,AizenmanWarzel2011,Froese2016}. Regular hyperbolic lattices retain loops but possess a microscopic lattice spacing and only a discrete point-group symmetry~\cite{Chen2024PRL,Li2024CommunPhys}. The present model instead has the local differential structure of a two-dimensional manifold, exact continuous rotations about the chosen origin, and a controlled $a\rightarrow\infty$ flat limit.

The distinction from continuum isotropic disorder is equally important. Altland \textit{et al.} derive a two-parameter renormalization flow for statistically isotropic bulk disorder on the hyperbolic plane and obtain a metal--insulator critical line~\cite{Altland2026}. Here the disorder statistics deliberately single out a radial origin: $V_r$ is common to every angle and the weaker $V_\perp$ is short-range correlated along physical arc length. The observable is the angular fragmentation of a prelocalized radial mode, not bulk conductance. Consequently, the logarithmic $r_*$ law is a shell-resolved crossover within an anisotropic ensemble and does not assert the absence of the isotropic critical line.

\begin{table*}[t]
\caption{Geometric and localization distinctions. Exponential growth is shared by all three geometries, while the final column identifies the ingredients specific to the present continuum mechanism. Citations mark representative evidence for the corresponding geometry or localization problem.}
\label{tab:comparison}
\small
\renewcommand{\arraystretch}{0.98}
\begin{tabular*}{\textwidth}{@{\extracolsep{\fill}}llll@{}}
\toprule
\parbox[t]{0.14\textwidth}{Property} &
\parbox[t]{0.21\textwidth}{Bethe lattice} &
\parbox[t]{0.25\textwidth}{Regular hyperbolic lattice} &
\parbox[t]{0.28\textwidth}{Continuous hyperbolic plane in this work}\\
\midrule
\parbox[t]{0.14\textwidth}{Local structure} &
\parbox[t]{0.21\textwidth}{Loopless graph with fixed coordination~\cite{AbouChacra1973,AizenmanWarzel2011}} &
\parbox[t]{0.25\textwidth}{Discrete polygonal graph with loops~\cite{Chen2024PRL,Li2024CommunPhys}} &
\parbox[t]{0.28\textwidth}{Smooth two-dimensional Riemannian manifold~\cite{Curtis2025PRL,Altland2026}}\\[2pt]
\addlinespace
\addlinespace
\parbox[t]{0.14\textwidth}{Rotations} &
\parbox[t]{0.21\textwidth}{No angular-momentum quantum number} &
\parbox[t]{0.25\textwidth}{Finite lattice point group} &
\parbox[t]{0.28\textwidth}{Exact $SO(2)$ symmetry and integer $m$ at $\varepsilon=0$}\\[2pt]
\addlinespace
\addlinespace
\parbox[t]{0.14\textwidth}{Ultraviolet scale} &
\parbox[t]{0.21\textwidth}{Fixed graph edge} &
\parbox[t]{0.25\textwidth}{Fixed lattice spacing} &
\parbox[t]{0.28\textwidth}{Continuum limit independent of numerical mesh}\\[2pt]
\addlinespace
\addlinespace
\parbox[t]{0.14\textwidth}{Transverse growth} &
\parbox[t]{0.21\textwidth}{Exponential in generation~\cite{AbouChacra1973,AizenmanWarzel2011}} &
\parbox[t]{0.25\textwidth}{Exponential in graph distance~\cite{Chen2024PRL,Li2024CommunPhys}} &
\parbox[t]{0.28\textwidth}{$L_-(r)=2\pi a\sinh(r/a)$: locally linear and asymptotically exponential}\\[2pt]
\addlinespace
\addlinespace
\parbox[t]{0.14\textwidth}{Flat limit} &
\parbox[t]{0.21\textwidth}{None} &
\parbox[t]{0.25\textwidth}{A continuum flat limit requires simultaneous lattice refinement} &
\parbox[t]{0.28\textwidth}{$a\rightarrow\infty$ gives $S_K(r)\rightarrow r$}\\[2pt]
\addlinespace
\addlinespace
\parbox[t]{0.14\textwidth}{Problem studied} &
\parbox[t]{0.21\textwidth}{Branching-assisted resonances} &
\parbox[t]{0.25\textwidth}{Generic-disorder transitions and mobility edges~\cite{Chen2024PRL,Li2024CommunPhys}} &
\parbox[t]{0.28\textwidth}{Weak angular disorder acting on a radially localized continuum mode}\\[2pt]
\addlinespace
\addlinespace
\parbox[t]{0.14\textwidth}{Specific result} &
\parbox[t]{0.21\textwidth}{Not applicable} &
\parbox[t]{0.25\textwidth}{No exact $SO(2)$ partial-wave decomposition} &
\parbox[t]{0.28\textwidth}{$m\leftrightarrow m+q$ and $r_*/a=\arsinh z$ with a controlled flat limit}\\
\bottomrule
\end{tabular*}
\end{table*}

If the projected perturbation is expanded as
\begin{equation}
 V_\alpha(\theta)=\sum_{q\in\mathbb Z}V_{\alpha q}\e^{\ii q\theta},
 \label{eq:angularfourier}
\end{equation}
then exact continuum angular momentum gives the selection rule $m\leftrightarrow m+q$. Neither a Bethe lattice nor a generic regular hyperbolic lattice has this exact partial-wave structure. More importantly, Eqs.~\eqref{eq:rhyper}--\eqref{eq:rflat} continuously connect the local Euclidean regime to the curvature-dominated logarithmic regime. The asymptotic logarithm has tree-like geometric ancestry, but the crossover function and its flat limit are properties of the continuous constant-curvature metric.

This distinction also clarifies the role of earlier curved-surface spin studies~\cite{ShimaSakaniwa2006JPA,ShimaSakaniwa2006JSTAT,BaekShimaKim2009,SakaniwaShima2009}. Those works showed how negative curvature modifies critical behavior and boundary scaling in interacting classical systems. Here curvature instead converts a one-dimensional localization length along a geodesic circle into a radial crossover distance for a continuum quantum mode. The common geometric ingredient is exponential circumference growth; the observables and disorder mechanism are different.

\section{Discussion}
\label{sec:discussion}

The analysis establishes three levels of physics that should not be conflated. The first is a symmetry-enforced circular profile in the exactly radial ensemble. The second is true radial Anderson localization, verified by a positive infinite-half-line Lyapunov exponent and absent in clean and periodic controls. The third is angular localization of the projected ring, whose onset depends on the ratio $L_K(r_0)/\xi_\theta$. Only the third level is curvature amplified.

The predicted logarithmic shell-breaking distance is robust to the detailed short-range angular disorder because weak one-dimensional localization generically gives $\xi_\theta\propto\varepsilon^{-2}$, while the hyperbolic circumference generically gives $L_-(r)\propto\e^{r/a}$. Disorder correlations alter the prefactor through $\widetilde C_\theta(2k_\theta)$ and can generate exceptional energies where the leading Born term vanishes, as in correlated one-dimensional systems~\cite{IzrailevKrokhin1999,ShimaNomuraNakayama2004}. Away from such engineered zeros, the coefficient changes but the logarithmic dependence remains.

The finite-ring calculations determine the projected angular localization length and operational coefficient $c_Q$. The direct polar-grid calculation then tests, for identical disorder realizations, whether this reduced description reproduces the corresponding sector of the full two-dimensional Hamiltonian. The small subspace mismatch and high retained weight quantify the projection error for narrow, spectrally isolated bulk shells across multiple radial-disorder realizations; they do not establish transport properties for generic isotropic disorder or for an expanding thermodynamic set of angular channels. A future study of that distinct problem should determine radial transport and the typical local density of states without imposing dominant ring correlations.

Potential realizations include wave systems in synthetic hyperbolic geometries, such as circuit, microwave, or photonic platforms~\cite{Kollar2019,Lenggenhager2022,Zhang2022,Huang2024}. A deliberately ring-correlated disorder profile would be required to access the solvable radial reference ensemble. The most direct observable would be the number of angular localization segments on shells of increasing geodesic radius, together with the predicted conversion from an arc localization length to a radial logarithmic scale.

\section{Conclusion}
\label{sec:conclusion}

We have formulated a continuum disorder problem that cleanly separates radial Anderson localization from the angular shape imposed by rotational symmetry. Exact partial-wave reduction gives independent one-dimensional random radial Hamiltonians. Clean and periodic controls confirm that a positive radial Lyapunov exponent is generated by disorder, and three outer boundary conditions confirm exponential insensitivity of bulk localized states to a finite disk edge.

Weak angle-dependent disorder projects each radial mode onto a one-dimensional random ring. Transfer matrices verify the quadratic angular Lyapunov exponent, and participation ratios collapse as a function of circumference divided by angular localization length. A direct full polar-grid calculation at matched energy validates the projection by unbiased subspace tracking over independent radial and angular disorder, while a longer weak-disorder extrapolation recovers the discrete Born coefficient. Constant-curvature geometry then converts this one-dimensional length into a shell-breaking radius. The result is algebraic in flat space, logarithmic on the hyperbolic plane, and may be unreachable on a positively curved hemisphere. The noncommuting weak-disorder and flat-curvature limits establish a singular crossover within the ring-correlated ensemble; no claim is made about the generic isotropic-disorder transition.

The phenomenon shares exponential-growth physics with trees and regular hyperbolic lattices but differs through exact continuum angular momentum, a smooth ultraviolet limit, and a controlled interpolation to the Euclidean plane. These features make curvature-amplified shell breaking a distinct continuum localization mechanism.

\appendix

\section{Derivation of the radial channel Hamiltonian}
\label{app:radialderivation}

For one angular harmonic, write $\psi=(2\pi)^{-1/2}S_K^{-1/2}u_m\e^{\ii m\theta}$. Direct differentiation gives
\begin{align}
 &\left(\partial_r^2+\frac{S_K'}{S_K}\partial_r\right)
 \frac{u_m}{\sqrt{S_K}} \nonumber\\
 &\qquad=\frac{1}{\sqrt{S_K}}
 \left[u_m''+\left(\frac{[S_K']^2}{4S_K^2}
 -\frac{S_K''}{2S_K}\right)u_m\right].
 \label{eq:radialidentity}
\end{align}
The angular derivative contributes $-m^2u_m/S_K^2$. Substitution into Eq.~\eqref{eq:Hfull} at $\varepsilon=0$ yields Eqs.~\eqref{eq:radialgeneral} and \eqref{eq:geometricpotential}. For $S_K=a\sinh(r/a)$, the identities $S_K''/S_K=a^{-2}$ and $[S_K'/S_K]^2=a^{-2}\coth^2(r/a)$ give Eq.~\eqref{eq:Uhyper}.

\section{Numerical procedures and reproducibility}
\label{app:numerics}

Numerical data and computational materials supporting Figs.~\ref{fig:radialcontrols}--\ref{fig:weaklog} are retained by the corresponding author and are not publicly deposited. Materials retained by the author, including machine-readable data, parameters, random-seed records, and source code, are available from the corresponding author upon reasonable request. The scripts used for the full-model and weak-asymptotic tests in Figs.~\ref{fig:projection} and \ref{fig:weaklog} use fixed pseudorandom seeds and no system entropy.

For Fig.~\ref{fig:radialcontrols}, exact $2\times2$ transfer matrices propagate Eq.~\eqref{eq:radialtest} through constant-potential cells. Panel (a) uses $k_r=1.2$, $\ell_r=0.5$, 120 realizations, and total length $2.0\times10^4$. Panel (b) plots $|\Gamma_R|$ for $R=10$--$10^4$ using the same $k_r$ and $\ell_r$: the clean potential is zero, the period-two control alternates between $\pm0.4$, and the random control is uniform in $[-0.4,0.4]$. The random curve is ensemble averaged; the clean and periodic values are deterministic.

For Fig.~\ref{fig:boundary}, sparse shift-invert diagonalization follows a localized state near the fixed target energy under Dirichlet, Neumann, and Robin outer conditions. State matching maximizes overlap on the bulk core, defined by excluding the outermost $2\xi_{\rm IPR}$ of the Dirichlet reference state. The plotted points are medians, and the spread bars are interquartile ranges. The conclusion uses the conservative criterion $d_b/\xi_{\rm IPR}\geq7$ rather than the fitted exponents, whose values depend on the operational IPR length.

For Fig.~\ref{fig:angular}, the angular Lyapunov exponent uses $1.2\times10^5$ transfer steps and 120 realizations. Ring eigenstates are obtained by sparse diagonalization for the circumferences and perturbation amplitudes specified in the parameter records retained by the author. Four states nearest $E_\theta=0.5$ are averaged for each disorder realization. For each $\varepsilon$, $N_*$ is obtained by linear interpolation between the two adjacent sizes bracketing $Q_{\rm sh}=1/2$; the reported $c_Q$ uncertainty is the standard deviation across the five perturbation strengths.

For Fig.~\ref{fig:ldos}, diagonal resolvent elements are evaluated at $E_\theta=0.5$ for periodic 4096-site chains and 48 realizations. The order of limits is not extrapolated; the figure is used only as a finite-size spectral illustration.

For Fig.~\ref{fig:projection}, the transformed Hamiltonian in Eq.~\eqref{eq:fullpolargrid} is assembled as a sparse matrix on the polar grid. Shift-invert diagonalization returns 12 candidate states near the target energy at each continuation step. The two-dimensional subspace is chosen only by overlap with the subspace at the preceding $\varepsilon$, beginning with the predetermined $\pm m_0$ subspace at $\varepsilon=0$. Six independent radial seeds and eight independent angular seeds per radial realization are used for the baseline curves. Grid, domain, and cumulative-sample convergence are evaluated separately from the state-resolved baseline data.

For Fig.~\ref{fig:weaklog}, the recurrence associated with Eq.~\eqref{eq:ringlattice} is normalized after every transfer step after an 8192-step burn-in. One hundred ninety-two realizations are propagated independently at each perturbation strength. The transfer length is selected from the Born estimate to give approximately 400 expected Lyapunov e-foldings, with the stated lower and upper bounds. Realization-level exponents, ensemble summaries, the weighted-fit covariance, and the curvature conversion are retained as supporting numerical records.

\section{Symbol glossary}
\label{app:symbols}

The geometric symbols are $K$ for Gaussian curvature, $a$ for curvature radius, $r$ and $\theta$ for geodesic polar coordinates, $S_K(r)$ for metric radius, and $L_K(r)$ for geodesic circumference. The quantum symbols are $M$ for particle mass, $\hbar$ for the reduced Planck constant, $E$ for energy, $m$ for angular momentum, $u_m$ for radial amplitude, and $U_{m,K}$ for geometric radial potential.

The disorder symbols are $V_r$ and $V_\perp$ for the radial and angular random potentials, $\varepsilon$ for the angular perturbation amplitude, $W_r$ for radial disorder width, $\ell_r$ for radial cell length, and $\ell_\theta$ for angular arc correlation length. The wave numbers are $k_r$ in the radial test problem and $k_\theta$ in the projected angular problem.

The localization symbols are $\gamma_r$ and $\xi_r$ for radial Lyapunov exponent and localization length, $\gamma_\theta$ and $\xi_\theta$ for their angular counterparts, $P_\theta$ for angular participation length, and $Q_{\rm sh}$ for the single-state normalized shell-extension indicator. The basis-independent two-state measure is $Q_{\rm sub}$. The projected metric symbols are $\beta_\alpha=\langle S_K^{-2}\rangle_\alpha$, $S_\alpha=\beta_\alpha^{-1/2}$, and $\delta_{\rm met}$ for its rms relative variation. The retained subspace weight and fidelity are $Z_{\rm sub}$ and $F_{\rm sub}$. The operational crossover symbols are $c_Q$ for the threshold coefficient, $N_*$ for crossover circumference, $r_*$ for crossover radius, and $z$ for the dimensionless curvature variable. The notation $\Vert A\Vert$ denotes the spectral norm of a matrix $A$ when Lyapunov exponents are computed from transfer matrices.

For boundary tests, $\bar r$ is the state center, $\xi_{\rm IPR}$ is the inverse-participation length, $d_b$ is distance from the outer boundary, and $d_{\min}$ is the smaller distance to either radial boundary. For spectral diagnostics, $\rho_{\rm av}$ and $\rho_{\rm typ}$ are average and typical local densities of states.

\section*{Data availability}
The numerical data and source code that support the findings of this article are not publicly available because a stable, versioned public archive has not been established for this study. The data and code are available from the corresponding author upon reasonable request.

\bibliography{HS_Anderson_Hyper_refs}

\end{document}